\documentclass[reprint,
 amsmath,amssymb,
 aps,prl,
]{revtex4-1}
\usepackage{graphicx}
\usepackage[normalem]{ulem}
\usepackage{dcolumn}
\usepackage{bm}
\usepackage{lipsum}
\usepackage[dvipsnames]{xcolor}

\begin{document}

\preprint{APS/123-QED}

\title{Liberation of slave modes inside  domain walls in multiferroic Cu-Cl boracite}

\author{Peng Chen}
\affiliation{Quantum Materials Theory, Istituto Italiano di Tecnologia,16163 Genova, Italy.}
\author{Sergey Artyukhin}
\affiliation{Quantum Materials Theory, Istituto Italiano di Tecnologia,16163 Genova, Italy.}

\date{\today}

\maketitle

{\bf Domain walls (DWs), the two-dimensional boundaries between symmetry equivalent ferroic domains,  are actively investigated due to their promise for novel logic and memory devices. Moreover, they can be easily created, erased and reshaped at a low energy cost due to their high mobility and large electrical conductivity. Most work so far has been focused on DWs in proper ferroelectrics, where the primary order parameter, ferroelectric polarization, interpolates between the values in the domains by either reducing to zero (in Ising-type DW) or rotating (Bloch type DW). Here we present a new member of DW family with a complex inner texture of slave order parameters inside the wall where the primary order parameter reduces to zero.
%
%
Our first-principles-derived model predicts the existence of monopolar and toroidal polarization patterns. The results enable large-scale phase field simulations of complex domain patterns in boracites and could inspire novel devices based on domain walls in improper ferroelectrics.}

{\bf Introduction}.
Boracites are among the first discovered multiferroic materials \cite{Ascher1964,Ascher1966}, which however remain a source of puzzling experimental data. They are improper ferroelectrics, meaning that the polarization is induced through an anharmonic coupling with the primary order parameter, a 6-component $X_5$ mode, representing antipolar ionic displacements. 
The same antipolar distortion are also generally observed in the famous perovskite ground state structure Pnma. However, the X5 modes are a secondary order parameters, slave to the in- and anti-phase rotation.
The parent structure, F$\bar{4}3m$, is chiral, and therefore the components of electric polarization and shear strains transform according to the same irreducible representations, resulting in improper ferroelasticity. High dimension of the primary order parameter and strong interactions with the secondary ones gives rise to a rich free energy landscape with many domains, and complex domain patterns, observed in experiments. This explains the growing interest to boracites within the emerging field of domain wall-based nanodevice design \cite{Catalan2012}. Extensive work has also been done on magnetism in boracites, and the interplay between structural and magnetic orders \cite{Toledano1985,Feng2018}.Presence of multiple interacting multicomponent orders positions boracites as an ideal playground for domain wall injection and manipulation \cite{McQuaid2017}.

In order to build a theoretical basis for understanding of these puzzling phenomena, here we determine the parameters of a first-principles-based Landau-type theory, describing interacting antipolar displacements, ferroelectric polarizatioin and strains in Cu-Cl boracite. Using the model, we simulate the domain wall structure and domain patterns.

\begin{figure}[b]
\includegraphics[width=\linewidth]{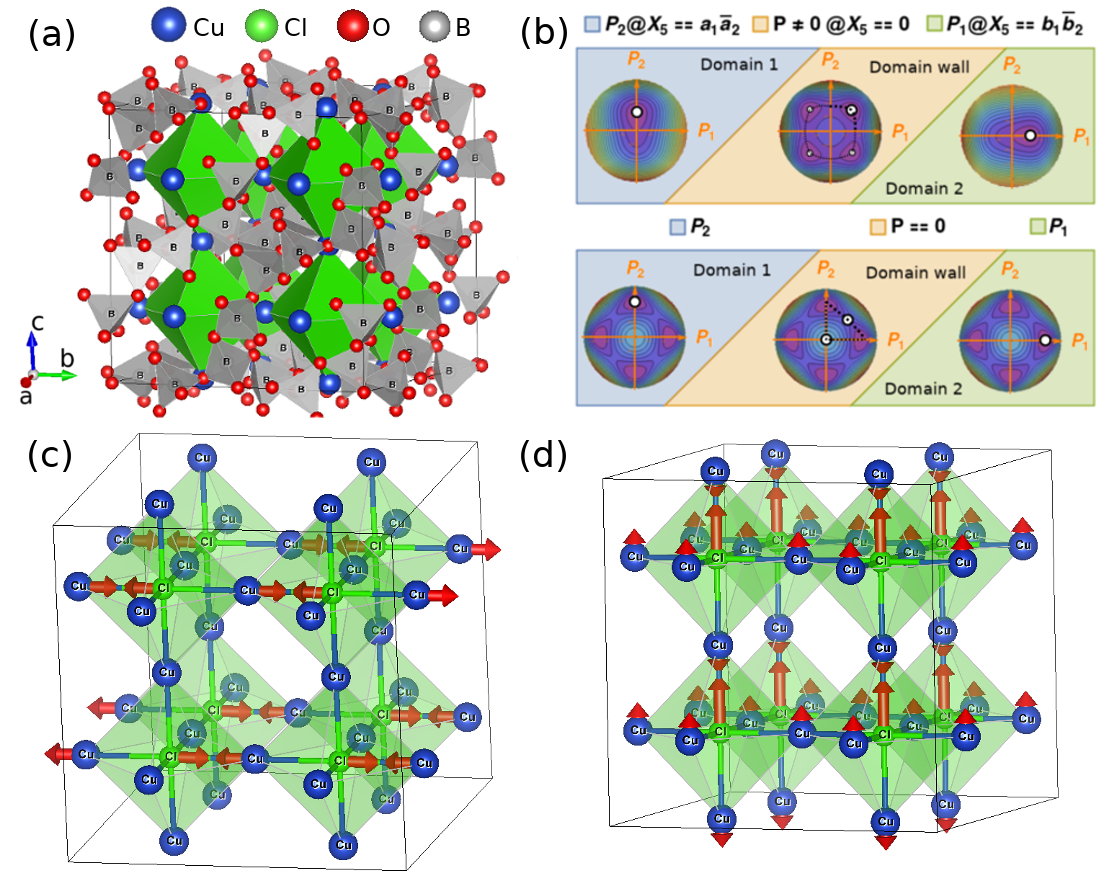}
\caption{\label{fig:struct}(a) Boracite parent structure F$\bar{4}$3c; it can be separated into (b) perovskite structure of XM$_3$ octahedra and (c) and clusters of BO$_4$ tetrahedra at the A site of ABO$_3$ structure; (d) antipolar distortion mode $c_1$; (e) polarization mode $P_3$.}
\end{figure}

\begin{figure*}[ht]
\includegraphics[width=\linewidth]{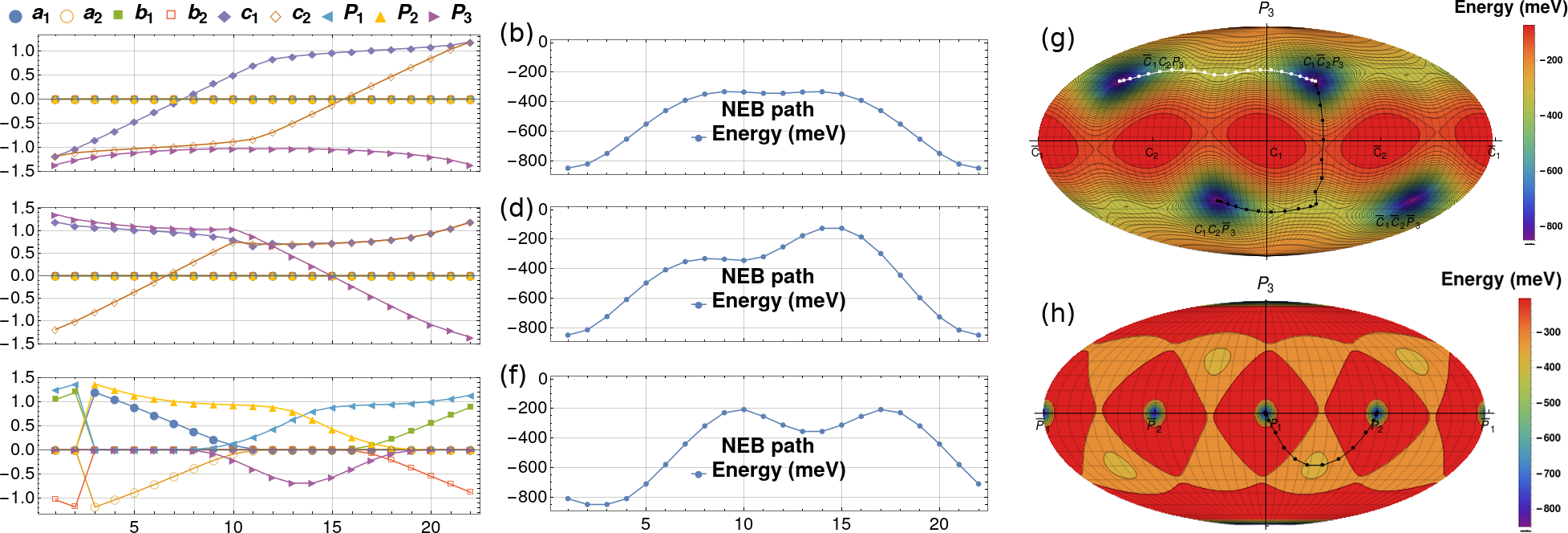}
\caption{\label{fig:NEB} A minimum energy path (MEP) from nudged elastic band (NEB) optimization. Energy along the MEP (a-c) and order parameters variations (d-f) across the domain wall between $\bar{c}_{1}\bar{c}_{2}\bar{P}_{3}$ and $c_{1}c_{2}\bar{P}_{3}$ in (a,d), 
$c_{1}\bar{c}_{2}P_{3}$ and $c_{1}c_{2}\bar{P}_{3}$ in (b,e), 
and $a_{1}\bar{a}_{2}P_{2}$ and $b_{1}\bar{b}_{2}P_{1}$ in (c,f); (g-h) Energy potential landscape on a Mollweide projection (see supplementary) with MEPs overlaid.}
\end{figure*}

{\bf Structure and Symmetry}.
Boracites are a crystal family with chemical formula M$_3$B$_7$O$_{13}$X, where M is a divalent metal and X is usually a halogen. 

The parent structure is F$\bar{4}$3c, and at lower temperatures boracites undergo phase transitions driven by the dominant distortion modes X$_5$ and $\Gamma_{4}$. At low temperatures Cu-Cl boracite adopts Pca2$_1$ structure. The group-subgroup analysis shows that the primary order parameter, corresponding to antipolar displacements, has six components $(a_1,a_2,b_1,b_2,c_1,c_2)$ and transforms according to the irreducible representation X$_5$, while polarization and shear strains transform as $\Gamma_{4}$. Components of $X_5$ with $a,b,c$ describe modulations with wavevectors along $a,b,c$ crystallographic directions. The distortions due to an antipolar mode $c_1$ and a polar mode P$_3$ are shown in Fig.~\ref{fig:struct}. $a_1$ and $b_1$ modes are obtained by acting on $c_1$ with the 3-fold rotation around the body diagonal. $P_1$ and $P_2$ modes are obtained from $P_3$ analogously. $\bar{4}$ operation transforms between $a_1$ and $a_2$ etc.
Only M and X ions are shown in Fig.~\ref{fig:struct}(d,e), as they contribute the most to the distortion, while the displacements of boron and oxygen ions are small. It is interesting that the boracite could be though of as a simple perovskite structure ABO$_3$ with corner-sharing ClCu$_6$ octahedra, while the clusters of BO$_4$ tetrahedra substitute A cations, and break the inversion symmetry, present in perovskites.
The absence of inversion symmetry in the parent structure of boracite allows exotic order parameter couplings, forbidden in typical perovskite-based ferroelectrics, such as BaTiO$_3$ and BiFeO$_3$. For example, boracite allows the coupling of polarization to X$_5$ modes, Eq.~\ref{eq:pc} $\gamma_{xp}P_{3}c_{1}c_{2}$, that gives rise to improper ferroelectricity, and a third order coupling among polarization components, Eq.~\ref{eq:ppp} $\gamma_pP_{1}P_{2}P_{3}$.
The complete list of invariants is presented in Table~S2.

{\bf DFT energetics}.
To evaluate the interaction parameters of the model, we performed DFT total energy and phonon calculations on a number of representative low energy structures of Cu-Cl boracite, listed in Table~S1.
The ground state structure is 794.2 meV/8f.u. below the parent one. From the DFT energies presented in Table~S1, we see that the structural chirality results in energy difference between the structures with the same $(c_1,c_2)$ but opposite $P_3$.
Similarly, an energy difference could also be seen between $P_1P_2P_3$ and $\bar{P}_1\bar{P}_2\bar{P}_3$ phases and between $\bar{a}_{1}\bar{b}_{1}\bar{c}_{1}$ and $a_{1}b_{1}c_{1}$ phases.
For the phase without $X_5$ and a ferroelectric polarization, the $a_{2}\bar{b}_{2}\bar{c}_{2}$ and $P_1P_2P_3$ are the low energy phases close to the ground state, which means that there is strong anisotropic coupling among $\Gamma_{4}$ polarization modes, and the same between $X_{5}$ antipolar displacement modes.
Due to the coupling between $\Gamma_{4}$ polarization and $X_{5}$ antipolar displacements, the lowest energy state is the $c_{1}c_{2}\bar{P}_{3}$ phase. 
The coupling between $X_{5}$ antipolar displacement and $\Gamma_{4}$ shear strain does not significantly reduce the energy, $\Delta E = -32.6$~meV/8f.u., from $c_{1}c_{2}$ to $c_{1}c_{2}e$ phase.
The $\Gamma_{4}$ polarization is another mode that couples to shear strain, and results in a similar energy gain, $\Delta E = -30.3$~meV/8f.u.

{\bf Extracting Landau model parameters}.
The coefficients are fitted from DFT and summarized in the Table~S2.
The negative $\gamma_{xc}$ and positive $\gamma_{xp}$ indicate that a positive primary order parameter $c_{1}c_{2}$ favours a positive shear strain and a negative polarization.
However, the negative $\gamma_{pc}$ tells that there is a strong competition between the positive shear strain and the negative polarization (induced by the same $X_5$, $c_{1}c_{2}$).
This interesting fact implies the competition between the two slave modes, the shear strain and polarization, due to the force from the master mode, $X_{5}$ antipolar displacements.
While the coupling to $X_5$ forces the amplitudes of polarization and strain modes, frustrating their coupling to each other, these competing interactions in boracite could lead to a peculiar behavior at a phase transition or inside a domain wall, where the primary $X_5$ order disappears. In addition, this competition stabilizes the zero value of $X_5$ mode in the metastable rhombohedral state via the coupling between $X_5$ and $\Gamma_4$. This is corroborated by the stable phonon spectrum in that state, shown in Fig.~S1
For example, if an external electric field is opposite to the polarization in one region, the energy may still be gained on its interaction with the shear strain, leading to anomalous ferroelectric DW motion, where a domain with polarization opposite to the electric field grows. This possibility is unfortunately precluded in Cu-Cl boracite, where the piezoelectric tensor component $d_{123}$ is not large enough. The negative coefficients ($\alpha_{xxXX}$, $\alpha_{xxyy}$, and $\alpha_{xxYY}$) of anisotropic terms indicate that the components of $X_{5}$ attract to each other. In addition, there's a strong attractive interaction $\gamma_x$ between $a_1,b_1$ and $c_1$, which explains the very low energy of $\bar{a}_{1}\bar{b}_{1}\bar{c}_{1}$ phase in Table~S1.
\begin{figure*}[t]
\includegraphics[width=\linewidth]{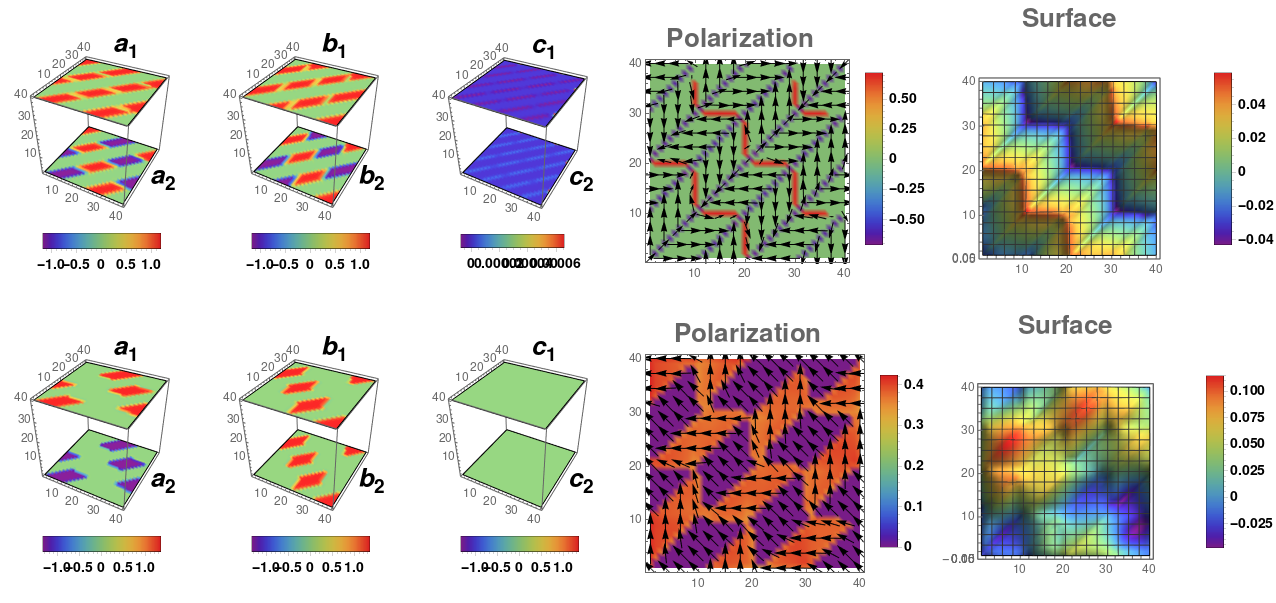}
\caption{\label{fig:Miura}The components of the $X_5$ mode for the Miura domain pattern with and without electric field. (a) $E=0$, (b) $E=E_0$, (c) $E=0$, (d) $E=E_0$; DW inside a DW is seen in panel (c) where the coloring of DW segments changes from yellow to blue. (e,f) Ferroelectric polarization texture corresponding to states represented in panels (a) and (d).}
\end{figure*}

{\bf Domain walls in Cu-Cl boracite}.
Now we turn to the structure of domain walls in boracites. The nudged elastic band (NEB) method is utilized to explore the minimum energy paths (MEP) connecting different domains, corresponding to the minima on the potential energy surface in Eq.~\ref{eq:one}.
 Fig.~\ref{fig:NEB}(a, d) shows the energy barrier and the order parameter variation for the NEB between $\bar{c}_{1}\bar{c}_{2}\bar{P}_{3}$ and $c_{1}c_{2}\bar{P}_{3}$ domains. The corresponding MEP is marked with a white curve in Fig.~\ref{fig:NEB}(g). Ferroelectric polarization does not change sign across the wall, and hardly changes along the path  (Fig.~\ref{fig:NEB}(d)), hence we call this 0$^\circ$ DW. The intermediate phase is a metastable local minimum $c_{1}\bar{c}_{2}\bar{P}_3$ (see Supplementary section ???).
 
 As for the DW between $c_{1}\bar{c}_{2}P_{3}$ and $c_{1}c_{2}\bar{P}_{3}$ phases, where polarization is reversed (180$^\circ$ ferroelectric DW), the MEP, shown with the black curve in Fig.~\ref{fig:NEB}(g), is asymmetric. MEP passes through a local minimum $c_{1}\bar{c}_{2}\bar{P}_{3}$ and a saddle point $c_{1}\bar{c}_{2}$. Note that an equivalent MEP through $c_{1}c_{2}$ and $c_{1}c_{2}P_{3}$ exists, potentially allowing to use an electric field to drive a hysteretic switching between the two. Under an external electric field, a segment of $c_1c_2$ wall will nucleate inside the $c_{1}\bar{c}_{2}\bar{P}_{3}$ wall, and the boundary between them would represent a 1D topological defect, that appears naturally in our simulations, e.g. shown in Fig.~\ref{fig:Miura}. Fig.~\ref{fig:NEB}(e) indicates that this is an Ising-type ferroelectric DW, so that $P_3$ changes, while other polarization components are zero.
 
 The most interesting is the ferroelectric 90$^\circ$ DW, whose MEP is shown in Fig.~\ref{fig:NEB}(h). It has a symmetric barrier. Suprisingly, the intermediate rhombohedral phase is $P_{1}P_{2}\bar{P}_{3}$, seen in Fig.~\ref{fig:NEB}(f), with $P_1=P_2=P_3$. Along the path, the X$_5$ components $a_{1}\bar{a}_{2}$ reduce to zero, however, the $b_{1}\bar{b}_{2}$ does not increase until the intermediate phase is reached (fig.~\ref{fig:NEB}(h)). Note that the primary order parameter is X$_5$, and the polarization is a slave order. When X$_5$ is present, the double well potential for the polarization is highly tilted, which is illustrated by contour plots inside domains in Fig.~\ref{fig:struct}(c). However, the NEB optimization indicates that X$_5$ is zero inside the 90$^\circ$ DW and, surprisingly, the slave polarization modes are liberated and become the primary modes inside the wall, which gives the DW a complex inner structure.

To the best of our knowledge, this special type of DW has never been reported, although it plays a key role in the formation of DW patterns and in DW motion discussed in the following section.

\begin{figure*}[ht!]
\includegraphics[width=\linewidth]{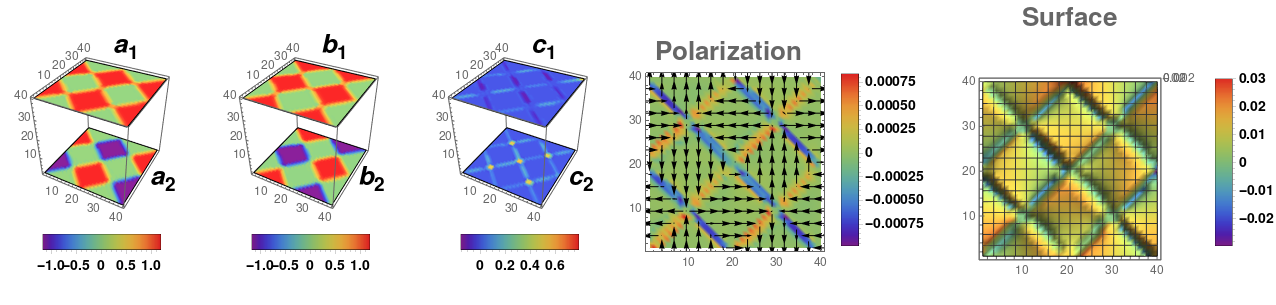}
\caption{\label{fig:toroidal}Miura, toroidal and ferrorotational polarization patterns in boracites.}
\end{figure*}

{\bf Miura patterns, monopolar and ferro-rotational polarization configurations}.

Paper can be folded into a famous Miura pattern, seen in Fig.~\ref{fig:Miura}, that conserves the area of the sheet. Similarly, $\epsilon_{xz}, \epsilon_{yz}$ strains in boracite tilt its surface but do not change its area, and therefore boracite domain adopt this pattern, since assembly of such sheared unit cells does not result in external surface tension.
In the Miura domain pattern, the vertical and horizontal DWs are 180$^\circ$ and all the diagonal DWs are 90$^\circ$ walls. As we see, DWs are in a rhombohedral phase, $P_1=P_2=P_3$.
When the electric field is applied, the domains with polarization along the field remain, while the domains with anti-parallel polarization are converted into the rhombohedral $P_1=P_2=P_3$ phase.
This suggests that domains in the Miura pattern with polarization along the field are protected, while others could be altered. Notably, the in-plane electric field here induces out-plane polarization and current.
These exotic phase transformations under electric filed could be utilized to implement memory read and write operations.
Taking advantage of the 90$^\circ$ DWs, ferro-rotational and monopolar polarization patterns, shown in Fig.~\ref{fig:toroidal} could also be injected in boracites via an application of a mechanical force.

{\bf Methods.}
The parameters of the Ginzburg-Landau model are extracted by fitting the total energies and phonon dispersion from first principle calculations to the model free energy,
\begin{widetext}
\begin{eqnarray}\label{eq:energy}
&&f = f_{x}+f_{p}+f_{c}+f_{xp}+f_{xc}+f_{pc}+G_X\nabla_iX_j\nabla_iX_j+G_P\nabla_iP_j\nabla_iP_j,\label{eq:one}\\
&&f_{x} = \alpha_{i}X_{i}^2+\alpha_{ij}X_{i}^2X_{j}^2+\alpha_{ijk}X_{i}^2X_{j}^2X_{k}^2+\gamma_{x}(a_{1}b_{1}c_{1}-a_{2}b_{2}c_{2}),\nonumber\\
&&f_{p} = \beta_{i}P_{i}^2 + \beta_{ij}P_{i}^2P_{j}^2 + \beta_{ijk}P_{i}^2P_{j}^2P_{k}^2+\gamma_{p}P_{1}P_{2}P_{3},\label{eq:ppp}\\
&&f_{c} = \frac{1}{2}\epsilon_{ij}C_{ijkl}\epsilon_{kl}+\gamma_{c}\epsilon_{13}\epsilon_{23}\epsilon_{12},\nonumber\\
&&f_{xp}= \gamma_{xp}(a_{1}a_{2}P_{2}+b_{1}b_{2}P_{1}+c_{1}c_{2}P_{z})+\eta_{xp}((a_{1}b_{2}c_{1}-a_{2}b_{1}c_{2})P_{1}+(a_{2}b_{1}c_{1}-a_{1}b_{2}c_{2})P_{2}+(a_{1}b_{1}c_{2}-a_{2}b_{2}c_{1})P_{3}),\label{eq:pc}\\
&&f_{xc} = \gamma_{xc}(a_{1}a_{2}\epsilon_{13}+b_{1}b_{2}\epsilon_{23}+c_{1}c_{2}\epsilon_{12})+\eta_{xc}((a_{1}b_{2}c_{1}-a_{2}b_{1}c_{2})\epsilon_{23}+(a_{2}b_{1}c_{1}-a_{1}b_{2}c_{2})\epsilon_{13}+(a_{1}b_{1}c_{2}-a_{2}b_{2}c_{1})\epsilon_{12}),\nonumber\\
&&f_{pc} = -\frac{1}{2}q_{ijkl}\epsilon_{ij}P_{k}P_{l}+\gamma_{pc}(P_{2}\epsilon_{13}+P_{1}\epsilon_{23}+P_{3}\epsilon_{12}),\nonumber
\end{eqnarray}
\end{widetext}
where the distortion modes $X_5\left( a_{1}, a_{2}, b_{1}, b_{2}, c_{2},c_{1}\right)$, polarization $\left(P_{1}, P_{2}, P_{3}\right)$, and unit cell displacements $\left(u_{1}, u_{2}, u_{3}\right)$, related to strain components via $\epsilon_{ij}=\frac{1}{2}(\partial_{j}u_{i}+\partial_{i}u_{j})$, are identified by group-subgroup analysis. The gradient terms in (\ref{eq:one}) penalize spatial variations of order parameters. $f_x$ and $f_p$ representa Mexican hat-like potentials for $X_5$ and $P$ modes; $f_c$ stands for the elastic energy, while $f_{xp}$ and $f_{xc}$ describe interactions between $X_5$ modes and ferroelectric polarizations and strains, respectively. $f_{pc}$ accounts for the energy due to electrostriction and a piezoelectric effect. Note that we have neglected $\gamma_{c}\epsilon_{23}\epsilon_{13}\epsilon_{12}$ term.
Although $\gamma_{c}$ drives instability of shear strains, calculations suggest the associated strain amplitude to be only $1/100$ of the ground state shear strain, and therefore we drop this term.

Density functional theory calculations \cite{DFT_1,DFT_2} are performed using VASP code \cite{vasp}, with projector-augmented wave formalism \cite{PAW1994,vasp_paw} and PBEsol exchange-correlation functional \cite{PBEsol}. The total energy calculations are performed with 500~eV energy cutoff and Monkhorst-Pack 3x3x3 k-mesh \cite{mp}. We use 192-atom supercells compatible with all the distortions discussed in this study. The representative structures were chosen in the following way. The symmetry inequivalent combinations of $X_5$ antipolar and $\Gamma_4$ polarization modes are frozen into the cubic structure and geometric optimization was performed to obtain all the structures, corresponding to all the energy minima and saddle points. The energies of those states and some others are reported in Table~S1. Then the training sets  for model fitting \cite{LInv2019} are generated by varying mode amplitudes with constant increments around each individual minimum and saddle point, until the energy change of tens of meV is reached. 

The elastic moduli and interatomic force constants are calculated using density functional perturbation theory \cite{Baroni2001}. Phonopy package \cite{phonopy} is used in the phonon spectrum calculations. Minimum energy paths were determined with the help of the nudged elastic band method \cite{neb} which can give the most energetically favorable intermediate configuration between the initial and final structures.



%

\end{document}